\documentstyle[12pt]{article}

\newcounter{subeq}
\newcommand{\eqnreset}{\setcounter{subeq}{0}}
\newcommand{\eqna}{\setcounter{subeq}{1}}
\newcommand{\eqnb}{\addtocounter{equation}{-1}{\setcounter{subeq}{2}}}
\newcommand{\eqnc}{\addtocounter{equation}{-1}{\setcounter{subeq}{3}}}

\newcommand{\Z}{\mbox{$Z^0$}}

\newcommand{\be}{\begin{equation}}
\newcommand{\ee}{\end{equation}}
\newcommand{\bea}{\begin{eqnarray}}
\newcommand{\eea}{\end{eqnarray}\eqnreset}
\begin{document}

%
%
\renewcommand{\theequation}{\arabic{equation}\alph{subeq}}
\renewcommand{\thefigure}{\arabic{figure}}
\renewcommand{\thetable}{\arabic{table}}
\renewcommand{\theenumi}{\roman{enumi}.)}
\renewcommand{\labelenumi}{\roman{enumi}.)}

\newcommand{\pngoes}{\raisebox{-3.0ex}{\mbox{
$\stackrel{\textstyle \longrightarrow}
{\stackrel{\scriptstyle | \vec{p}_i | >\!> m}
{\scriptstyle n\, {\rm large} } } \ $}}}

\def\be{\begin{equation}}
\def\ee{\end{equation}}
\def\Z{$Z^{\circ}$ }
\newcommand{\si}{$\sin^2 \Theta_W$}
\def \e{$|K_{L, S} (0) > $}
\def \h{$ H_{\rm eff}$}

\rightline{}
\title{CP and CPT Violation:  Status and Prospects\thanks
{Invited talk given at the 1993 Rencontre de Physique de la Valle
d'Aoste, La Thuile, Italy, March 1993}}
\author{R .D.~Peccei \\
 {\footnotesize {Department of Physics, University of
   California, Los Angeles, CA 90024 }}}
\maketitle
\begin{abstract}

I review the status of CP violating phenomena and CPT tests in the
neutral Kaon system.  Comparisons of present data with the expectations
of the Cabibbo Kobayashi Maskawa model are presented, with particular
emphasis being focused on the role of the theoretical uncertainties in
this comparison.  In addition, novel tests of CPT at DAFNE and Fermilab
are briefly discussed.
\end{abstract}

\addtocounter{page}{-1}
\newpage
\section{Introduction}

The study of CP violating phenomena is by now a mature phenomena in
particle physics, as the subject matter is nearly 30 years old!
Unfortunately, even after having performed very sophisticated
experiments, our information on CP violation is still very limited.
Basically, at present, we have:

\begin{enumerate}
\item Some positive evidence for the violation of $CP$ in the $K^0 -
{\bar{K}}^0$ complex, as a result of measuring non vanishing values for
the parameters $\eta_{+ -}, \eta_{00}$ and $A_{K_{L}}$.
\item Some bounds on the electric dipole moments of various particles
(e.g., for the neutron we know that $d_n < 1.2 \times 10^{-25}$~ecm,
while for the electron present data gives $d_e = (-0.3 \pm 0.8) \times
10^{-26}$ ecm \protect\cite{PDG})
\end{enumerate}

Our theoretical understanding of CP violation is marginally better.  In
the standard model of the electroweak interactions there is a paradigm -
the CKM paradigm - which accounts for CP violation.  According to this
paradigm, CP is violated because of the presence of a complex phase in
the mixing matrix for quarks - the so, called, Cabibbo Kobayashi Maskawa
matrix \cite{CKM}-with this phase originating in the symmetry breaking
sector of the theory.  However, at present this paradigm is only
qualitatively, but not quantitatively, confirmed by the data.
Furthermore, serious theoretical uncertainties plague this comparison.

The situation is perhaps better regarding CPT tests.  First of all, CPT
invariance, is expected to hold on the basis of deep theoretical
principles.  Any theory which is described  by a local, Lorentz
invariant Lagrangian, and in which there is a normal connection between
the spin and the statistics obeyed by the particle excitations, respects
CPT exactly \cite{CPT}.  Experimentally, no significant violations of
CPT exist.  Nevertheless, even here, the most accurate present tests of
CPT which are carried out in the neutral Kaon complex are not totally
unambiguous and could mask some possible CPT violations \cite{Dib}.

In this talk, I would like to review the status of CP violating
phenomena and of the present tests of CPT in the neutral Kaon system.  After
this brief review, I shall focus on two special topics:
\begin{enumerate}
\item How much do theoretical uncertainties influence the comparison of
data with the expectation of the CKM model.
\item What novel tests of CPT can be expected from the Phi factory now
being built at Frascati, as well as from a recently proposed experiment
to measure the antiproton lifetime at Fermilab.
\end{enumerate}
\section{Status of CP Violation and CPT Tests in the Neutral Kaon System}

To study CP violation and CPT tests in the $K^0 - {\bar{K}}^0$ complex
it has been traditional to describe this system by an effective $2
\times 2$ Hamiltonian \cite{Lee}
\be
H_{\rm eff} = M - \frac{i}{2} \Gamma~~,
\ee
where both the mass matrix $M$ and the decay matrix $\Gamma$ are
Hermitian matrices.  The time evolution of the system is then described
by the Schroedinger equation
\be
i \frac{\partial}{\partial t} \left(\frac{K^0}{K^0}\right) = H_{\rm eff}
\left(\frac{K^0}{K^0} \right)
\ee
It is possible to imagine \cite{Ellis}, however, that CPT violating
phenomena are connected with violations of quantum mechanics.  In this
case, clearly, the above simple Schroedinger equation is no longer
adequate and a more general analysis is required.  In what
follows, I will
not consider this more radical suggestion and describe possible
CPT violating phenomena within the usual 2-state formalism of quantum
mechanics.

The physical eigenstates, describing the $K_L$ and $K_S$  states, are
obtained by diagonalizing the above $2 \times 2$ Schroedinger equation
and these states evolve in time in the expected fashion:
\be
\vert K_{L, S}~ (t) > = e^{-im_{L,S} t} e^{-\frac{1}{2}\Gamma_{L,S}
t}~|~ {K_L,_S} (0) > ~~.
\ee
The eigenstates \e~ are linear combinations of $ {\vert{K}}^0 >$ and $\vert
{\bar{K}}^0 >$. If CP and CPT are conserved by $H_{\rm eff},$
the physical eigenstates
are CP eigenstates, otherwise they are not.
Diagonalizing the
Schroedinger equation one finds in the general case, when there are no CP
or CPT restrictions:
\bea
\vert K_S > &\simeq& \frac{1}{\sqrt{2}} \left\{ (1 + \epsilon_K +
\delta_K)
 \vert K^0 > + ( 1 - \epsilon_K - \delta_K ) \vert {\bar{K}}^0 >
\right\} \nonumber \\
\vert K_L > &\simeq& \frac{1}{\sqrt{2}} \left\{ ( 1 + \epsilon_K -
\delta_K) \vert K^0 > - ( 1 - \epsilon_K + \delta_K ) \vert {\bar{K}}_0 >
\right\}~~.
\eea
Here $\epsilon_K$ is a parameter that details the amount of CP violation
in $H_{\rm eff}$,
\be
\epsilon_K = e^{i \phi_{sw}} \frac{[ - Im M_{12} + \frac{i}{2} Im
\Gamma_{12}]}{\sqrt{2} \Delta m}~~,
\ee
while $\delta_K$ details possible CPT violation in \h:
\be
\delta_K = ie^{i \phi_{sw}} \frac{[(M_{11} - M_{22}) - \frac{i}{2}
(\Gamma_{11} - \Gamma_{22})]}{2\sqrt{2} \Delta m}~~.
\ee
In the above $\Delta m = m_L - m_S$ is the mass difference between the
eigenstates and $\phi_{sw}$ is a kinematical phase related to the ratio
of this mass difference to the difference in the $K_S$ and $K_L$ widths
\be
\phi_{sw} = \tan^{-1} \frac{2 \Delta m}{\Gamma_S - \Gamma_L} \simeq
45^0~~.
\ee

CP and CPT violations in the
$K^0 - {\bar{K}}^0$ system, besides through
$\epsilon_K$ and $\delta_K$, can enter also directly in the decay
amplitudes.  Essentially CP violation introduces further phases in these
amplitudes, while CPT violation is described by introducing further
amplitudes -  since particle and antiparticle decay amplitudes are then no
longer related. For semileptonic decays and for $K$ decays into $2 \pi$,
which will be of interest here, one can write \cite{Buchanan}
\bea
A (K^0 \to \pi^- \ell^+ \nu_{\ell}) = a + b~~;~~ & A({\bar{K}}^0 \to \pi^+
\ell^- {\bar{\nu}}_{\ell}) = a^{\ast} - b^{\ast} \nonumber  \\
 & \nonumber \\
A (K^0 \to 2 \pi; I) = (A_I + B_I) e^{i \delta_I}~~;~~ & A ({\bar{K}}^0
\to 2 \pi; I) = (A^{\ast}_I - B^{\ast}_I ) e^{i \delta_I}
\eea
In the above, $\delta_I$ is the usual $\pi \pi$ rescattering phase for
states in isospin $I~~(I = 0,2)$.  Having nonvanishing $b$ and
$B_I$ amplitudes signals CPT violation, while any CP violation makes
the $a$ and $A_I$ amplitudes complex.  Of course, observable effects in
the $K^0 - {\bar{K}}^0$ complex will measure a mixture of CP (and CPT)
violating decay and mixing parameters.

In the neutral Kaon system one has, at present, 5 measurements related
to CP and CPT violation. These involve
two (complex) amplitude ratios $\eta_{+
-}$ and $\eta_{00}$ and the semileptonic asymmetry $A_{K_{L}}$:

\bea \eqna \eta_{+ -} &=& \frac{A (K_L \to \pi^+ \pi^- )}
{A (K_S \to \pi^+ \pi^-)} =
\vert \eta_{+ -}\vert e^{i \phi_{+ -}} = \epsilon +
\epsilon^{\prime}\\ \eqnb
\eta_{00} &=& \frac{A (K_L \to \pi^0 \pi^0 )}{A (K_S \to \pi^0 \pi^0)} =
\vert \eta_{00}\vert e^{i \phi_{00}} = \epsilon -2
\epsilon^{\prime}\\ \eqnc
A_{K_{L}} &=& \frac{\Gamma (K_L \to \pi^- \ell^+ \nu_e ) - \Gamma(K_L \to
\pi^+ \ell^-{\bar{\nu}}_e)}{\Gamma (K_L \to \pi^- \ell^+ \nu_e) + \Gamma(K_L
\to \pi^+ \ell^- {\bar{\nu}}_e)}~~.
\eea
Experimentally one finds, to a good approximation, that
\begin{description}
\item{i)} $\vert \eta_{+ -} \vert \simeq \vert \eta_{00} \vert$
\end{description}
and
\begin{description}
 \item{ii)}  $A_{K_{L}} \simeq 2 Re~ \eta_{+-}$ and $\phi_{+-} \simeq
\phi_{00}$~~.
\end{description}
The first result shows that CP (or CPT) violation is essentially due to
mixing, since only the $\epsilon$ parameter, related to $\Delta S = 2$
processes, is important.  The second results, as we shall see, indicate
that experiments are consistent with CP violation, but CPT conservation.

For the study of CP violation and for comparing with the CKM paradigm it
is important to know if $\epsilon^{\prime} \neq 0$.  After all,
$\epsilon^{\prime}$ is a $\Delta S = 1$ parameter and, if CP violation
arises from quark mixing, that is precisely where one would expect to see
an effect.  Unfortunately, here the present experimental evidence is
conflicting.  One has information on $R e~ \epsilon^{\prime}/\epsilon$
from the ratio of rates, while $Im~ \epsilon^{\prime}/\epsilon$ can be
gleaned from the phase difference between  $\phi_{+ -}$ and $\phi_{00}$:
\be
\frac{\vert \eta_{+ -} \vert^2}{\vert \eta_{00} \vert^2} \simeq 1 + 6 Re
\frac{\epsilon^{\prime}}{\epsilon}~~;~~~ \phi_{+-} -  \phi_{00} \simeq 3
Im \frac{\epsilon^{\prime}}{\epsilon}~~.
\ee
Experimentally the most recent results obtained by the NA31 \cite{NA31}
and the E731\cite{E731} collaborations are

\bea Re \frac{\epsilon^{\prime}}{\epsilon} = \left\{\begin{array}{cc}
(23 \pm 7) \times 10^{-4}&\mbox{NA31}\\
(7.4 \pm 5.9) \times 10^{-4}&\mbox{E731}
\end{array}\right.
\eea
and
\bea \phi_{+-} - \phi_{00} = \left\{\begin{array}{cc}
(-0.2 \pm 2.6 \pm 1.2)^0 &\mbox{NA31} \\
( 1.6 \pm 1.0 \pm 0.7)^0 & \mbox{E731}
\end{array}\right.
\eea

I will return to discuss the CKM expectation for
$\epsilon^{\prime}/\epsilon$, after discussing how the present data in
the $K^0 - {\bar{K}}^0$  complex constrains CPT violating parameters.  I
note here only that because these CPT tests are somewhat less stringent
numerically, and $\epsilon^{\prime}$ is small compared to
$\epsilon$, it suffices for these tests to assume simply that $\epsilon
\simeq \eta_{+-}$, both in magnitude and phase.  The first test of CPT
arises from a comparison of the measured value of the semileptonic
asymmetry $A_{K_{L}}$ and $Re~ \epsilon$.  Straightforward calculations
\cite{Dib} yield the following expressions for $A_{K_{L}}$ and
$\epsilon$, to first order in small quantities,
\bea
A_{K_{L}} &=&  2 Re~ \epsilon_K + \biggl[2 \frac{Re~b}{Re~a} - 2 Re~
\delta_K\biggr ]
\nonumber \\
\epsilon &=& \epsilon_K + i \frac{Im A_0}{Re A_0} + \biggl[ \frac{Re B_0}
{Re A_0} - \delta_K \biggr] \nonumber \\
&\simeq& \frac{1}{\sqrt{2}} \left\{ - \frac{Im M_{12}}{\Delta m} +
\frac{Im A_0}{Re A_0} \right\} e^{i \phi_{sw}} + \frac{i}{\sqrt{2}}
\biggl[\frac{M_{22} - M_{11}}{2\Delta m} - \frac{Re B_0}{Re A_0}\biggr] e^{i
\phi_{sw}}~~.\nonumber \\
\eea
In the above, all quantities which violate CPT are enclosed in square
brackets.  The second expression for $\epsilon$ arises from saturating
the $2 \times 2$ width matrix $\Gamma$ by the $2 \pi, I=0$ states -  which
is an extremely good approximation\cite{Cronin}.
In this approximation the CP and
CPT violating components of $\epsilon$ are $90^0$ out of phase\cite{Cronin}

\be \epsilon =  \epsilon_{c \not{p}} e^{i \phi_{sw}} + \epsilon_{c\not{p}t}
e^{i(\phi_{sw} + \frac{\pi}{2})}~~,
\ee
with the CP violating component having the superweak phase $\phi_{sw}
\simeq 45^0$.  Using the $PDG$ \cite{PDG} values for $A_{K_{L}}$
and $\eta_{+-}$, one finds for the CPT violating amplitude difference
\be
\frac{Re B_0}{Re A_0} - \frac{Re~ b}{Re~ a} = Re~ \epsilon - \frac{1}{2}
A_{K_{L}} = ( - 0.6 \pm 0.7) \times 10^{-4}~~~{\rm PDG}
\ee

The difference between $\phi_{\epsilon}$, the phase of $\epsilon$
(essentially the phase of $\eta_{+-}, \phi_{+-})$, and the superweak
phase $\phi_{sw}$ provides a second test of CPT.  Using again $PDG$
values\cite{PDG}, there is about a $ 2 \sigma$ difference
between $\phi_{\epsilon} \simeq \phi_{+-} = (46 \pm 1.2)^0$
\pagebreak
\begin{figure}[t]
\vspace*{6cm}
\caption[ ]{Plot of $\epsilon$ in the complex plane.  Note that the
difference (if any) between $\phi_{\epsilon}$ and $\phi_{sw}$ is grossly
exaggerated.}
\end{figure}
and the
superweak phase $\phi_{sw} = (43.73 \pm 0.14)^0$.  As is clear from
Figure 1, one has
\be \tan (\phi_{\epsilon} - \phi_{sw}) =
\frac{\epsilon_{c\not{p}t}}{\epsilon_{c \not{p}}} =
(4.0 \pm 2.2) \times 10^{-2}~~~PDG~~.
\ee
Drawing any conclusion about a possible violation of CPT from the above
is quite premature.  Indeed, the actual value of $\phi_{+-}$ (and thus
of $\phi_{\epsilon}$) obtained from experiment is quite sensitively
dependent on the values of $\Delta m$ and, to a lesser extent, of
$\Gamma_S$ one uses.  This is very clear from the recent analysis
presented by the $E731$ collaboration\cite{Phase} and it is worthwhile
to repeat their arguments here.

$E731$ first fits the time evolution of their signal after the regenerator
\cite{Phase}
\be
\frac{dN}{dz} = \vert \rho_{\rm reg} e^{-\frac{z \Gamma_S}{2 \gamma}}
e^{i \frac{\Delta m z}{\gamma}} + \eta_{+-} e^{ - \frac{z \Gamma_L}{2
\gamma}} \vert^2
\ee
to obtain $\Delta m$ (and $\Gamma_S)$, keeping $\phi_{+-}$ fixed at
$\phi_{+-} = \phi_{sw}$.  When they do this, they obtain a value
for $\Delta m$ about 2 $\sigma$ below the value of $\Delta m$ quoted in
the PDG \cite{PDG}. [$\Delta m = (0.5286 \pm 0.0028)\times
10^{10}~{\rm sec}^{-1}$ \cite{Phase} versus $\Delta m = (0.5351 \pm 0.0024)
\times 10^{10}{\rm sec}^{-1}$~~\cite{PDG}].  Next, they let both $\Delta m$
and $\phi_{+-}$ float, getting a value for $\Delta m$ consistent with the
value they obtained earlier (but with bigger errors) and find
\be
\phi_{+-} = (42.2 \pm 1.4)^0~~~{\rm E731}~~,
\ee
a value entirely in agreement with $\phi_{sw}$ \footnote{Because of
their somewhat smaller $\Delta m$ value, the $E731$ collaboration has
also a somewhat smaller value for $\phi_{sw}$ than the value one would
infer from the PDG.  They find,\cite{Phase} $\phi_{sw} = (43.4 \pm
0.2)^0$.}.  Perhaps most importantly, if one uses the new E731 $\Delta m
$ value to renormalize some of the older experiments, whose values for
$\phi_{+-}$ were used to get the PDG value for $\phi_{+-}$, one gets a
considerable shift {\bf downward} for $\phi_{+-}$.  This is summarized
in Table 1, adapted from\cite{Phase}.
Combining these new values for $\phi_{+-}$ with the value of $\phi_{+-}$
obtained by E731\cite{Phase}, yields a new average value
\be
\phi_{+-} = (42.8 \pm 1.1)^0~~~{\rm New}~~.
\ee
Using this value (and the value of $\phi_{sw}$ from\cite{Phase}) gives
for the two CPT tests discussed above the results:
\bea
\frac{Re B_0}{Re A_0} &-& \frac{Re~ b}{Re a} = (0.3 \pm 0.7) \times
10^{-4}~~~{\rm New} \nonumber \\
\frac{\epsilon_{c \not{p}t}}{\epsilon_{c\not{p}}} &=& (-1.1 \pm 2.0)
\times 10^{-2}~~~  {\rm New}~~,
\eea
which are perfectly consistent with CPT conservation.
\begin{table}[h]
\caption{Old and New Values for\protect$\phi_{+-}$}
\vspace*{.2cm}
\begin{center}
\begin{tabular}{|l|l|l|} \hline
Experiment & $(\phi_{+-})~ {\rm Old} $ & $(\phi_{+-}) ~{\rm New}$ \\
\hline
Geweniger et al\protect\cite{G} & $(49.4 \pm 1.0)^0$ & $(43.0 \pm 1.0)^0$ \\
Carithers et al\protect\cite{C} & $(45.5 \pm 2.8)^0$ & $(44.0 ) \pm 2.8)^0$ \\
NA31\protect\cite{NA31} & $(46.9 \pm 1.6)^0$ & $(43.4 \pm 1.6)^0$ \\ \hline
\end{tabular}
\end{center}
\end{table}

I should remark that, since $\phi_{sw} \simeq 45^0$, one can simply
relate some of the other CPT violating parameters to
$\epsilon_{c\not{p}t}$. One finds \cite{Dib}
\be
Im \delta_K \simeq \frac{Re B_0}{Re A_0} - Re \delta_K \simeq
\frac{1}{\sqrt{2}}{\epsilon_{ c\not{p}t}}~~.
\ee
It is straightforward to show from the above, using the $2 \pi~ I = 0 $
approximation in $\Gamma$, that the bounds on $\epsilon_{c\not{p}t}$ obtained
imply the following bound for the diagonal parameters in $M$ and
$\Gamma$:
\bea
\frac{(M_{11} - M_{22}) + \frac{1}{2} (\Gamma_{11} - \Gamma_{22})}{4
\Delta m} = \left\{\begin{array}{cc}
(0.64 \pm 0.36) \times 10^{-4}~~~& \mbox{\rm PDG }\\
(-0.18 \pm 0.32)\times 10^{-4}~~~& \mbox{\rm New}
\end{array}\right.
\eea

There is a third test of CPT which is possible with present data.  This
test uses the fact that also in $\epsilon^{\prime}$ terms that violate
CPT are $90^0$ out of phase compared to terms that violate CP.  One
finds, \cite{Dib}
\be
\epsilon^{\prime} = \frac{Re A_2}{\sqrt{2} Re A_0} e^{i (\delta_2 -
\delta_0 + \frac{\pi}{2})} \left\{ \frac{Im A_2}{Re
A_2} - \frac{Im A_0}{Re A_0} + i \left[ \frac{Re B_0}{Re A_0} -
\frac{Re B_2}{Re A_2} \right] \right\}~~,
\ee
where I used again the convention that CPT violating terms are put in
between square brackets.  The phase of $\epsilon^{\prime}$ depends on
the $\pi \pi$ phase shifts $\delta_0$ and $\delta_2$ and, remarkably,
turns out also to be near $45^0$.  Indeed, recent analyses give

\bea
\phi_{\epsilon^{\prime}} = \delta_2 - \delta_0 + \frac{\pi}{2} =
 \left\{\begin{array}{cc}
(45 \pm 6)^0~~~&\protect\cite{GM}\\
(43 \pm 6)^0~~~&\protect\cite{Ochs}
\end{array}\right.
\eea
Because of this circumstance, to a very good approximation, $Im~
\epsilon^{\prime}/\epsilon$ will measure only the CPT violating
combination of parameters entering in $\epsilon^{\prime}$:
\be
\epsilon^{\prime}_{c\not{p}t} = \frac{Re B_0}{Re A_0} - \frac{Re
B_2}{Re A_2}~~.
\ee
Using experimental information on the magnitude of the $K^0 \to 2 \pi$
amplitudes and of $\vert \epsilon \vert$, one can write
\be
Im \frac{\epsilon^{\prime}}{\epsilon} \simeq \left( \frac{Re
A_2}{\sqrt{2} Re A_0 \vert \epsilon \vert} \right) \epsilon^{\prime}_{c
\not{p}t} \simeq 14 \epsilon^{\prime}_{c \not{p} t}~~.
\ee
A value for $\epsilon^{\prime}_{c \not{p}t}$ then follows from the
experimental values for the phase difference $\phi_{+-} - \phi_{00}
\simeq 3 Im~ \epsilon^{\prime}/\epsilon$.  Using data from NA31 and E731
gives the third CPT test:
\bea
\epsilon^{\prime}_{c\not{p}t} = \frac{Re B_0}{Re A_0} - \frac{Re B_2}{Re A_2}
= \left\{\begin{array}{cc}
(-0.8 \pm 11.9) \times 10^{-4}~~~&\protect\cite{NA31}\\
(6.7 \pm 5.1)\times 10^{-4}~~~&\protect\cite{E731}
\end{array}\right.
\eea

Before concluding this section, it is useful to make two remarks
concerning tests of CPT in the neutral Kaon complex.  First, as the
results presented show, CPT is tested in the ratio of CPT violating to
CPT conserving amplitudes (or in the ratio of diagonal element
differences of $M$ and $\Gamma$ to $\Delta m$) at the $10^{-4}$ level.
Improving the experimental accuracy of these CPT tests much beyond this
level appears very difficult to do.  However, and this is the second
remark I wanted to make, {\bf all} present day tests involve differences
of CPT violating quantities \cite{Dib}.  Thus, although unlikely, one
could imagine that the null tests of CPT violation obtained so far
result from an accidental cancellation!  As I will discuss in some
detail below, a $\Phi$ factory like DAFNE is ideally suited to test this
notion.  Before doing so, however, I want to discuss how well data in
the neutral Kaon system, along with some information from $B$ decays,
tests the CKM paradigm.
\section{Comparison with the CKM model - the role of theoretical
uncertainties}

The measurements of $\epsilon$ and $\epsilon^{\prime}$, in principle,
should provide confirmation of the CKM paradigm.  In practice, however,
one is hampered by various theoretical uncertainties.  To discuss this
comparison, it is useful to adopt the Wolfenstein parameterization
\cite{Wolf} of the mixing matrix $V_{CKM}$, in which one expands the
three mixing angles $\theta_1, \theta_2$ and $\theta_3$ in terms of
powers of the Cabibbo angle.  One write for these angles, in the
parameterization of $V_{CKM}$ adopted by the PDG \cite{PDG}, $\sin
\theta_1 = \lambda~~;~~ \sin \theta_2 = A \lambda^2~~;~~ \sin \theta_3 =
A \sigma \lambda^3$.   Here $\lambda = \sin \theta_C \simeq 0.22$ is the
sine of the Cabibbo angle and the parameters $A$ and $\sigma$ - which
turn out to be of $0 (1)$ - need to be fixed by experiment.  To $0
(\lambda^4)$ then one can write $V_{CKM}$ as:

\bea V_{CKM} \left| \begin{array}{ccc}
V_{ud} & V_{us} & V_{ub}  \\
V_{cd} & V_{cs} & V_{cb}  \\
V_{td} & V_{ts} & V_{tb} \end{array} \right| =\left| \begin{array}{ccc}
1 - \frac{\lambda^2}{2}  & \lambda& A \sigma \lambda^3 e^{- i \delta}  \\
- \lambda & 1 - \frac{\lambda^2}{2} & A \lambda^2  \\
A \lambda^3 (1 - \sigma e^{i \delta}) & -A \lambda^2 &
1 \end{array} \right|~~. \eea
The phase $\delta$ in the above is the phase responsible for CP
violation in the CKM paradigm.  Many authors, including Wolfenstein
\cite{Wolf}, instead of using the parameters $\sigma$ and $\delta$ in
$V_{CKM}$ use two other parameters $\rho$ and $\eta$, with $\eta$ being
connected to CP violation.  One has
\be
\sigma e^{-i \delta} = \rho- i \eta
\ee
or
\be
\rho = \sigma \cos \delta~~;~~~\eta = \sigma \sin \delta~~.
\ee

To extract the phase $\delta$ (or the parameter $\eta$) from the
measured values of $\epsilon$ and $\epsilon^{\prime}$, one needs
to know the value of the matrix elements of certain weak operators
involving quark fields between hadronic states.  Besides these hadronic
matrix elements, one also needs to know the value of the $A$ and $\sigma
$ (or $A$ and $\rho$) parameters in the CKM matrix, as well as a value
for the top quark mass, $m_t$.  All of these quantities are known with a
varying degree of accuracy and, as a result, the tests of the CKM
paradigm through the measurements of $\epsilon$ and $\epsilon^{\prime}$
are more qualitatative than quantitative.  Nevertheless, it is worthwhile
to trace the sources of the uncertainties and try to see what the
implications of these uncertainties are for testing the CKM paradigm.

The uncertainty in the parameters $A$ in $V_{CKM}$ is essentially that
of the $V_{cb}$ matrix element of this matrix.  The parameter $\sigma$
or $\sqrt{\rho^2 + \eta^2}$, on the other hand, depends on how well one
can determine the ratio of $V_{ub}$ to $V_{cb}$ in $V_{CKM}$.  Although
$\delta$~( or $\eta$) reflects the presence of CP violation,
constraints on this phase (or on this parameter) can also be inferred
from the magnitude of $\vert V_{td} \vert$.  This matrix elements of
$V_{CKM}$ can be deduced from the experimentally measured rate for $B -
\bar{B}$ mixing. However, also here to extract $\vert  V_{td} \vert$ one
needs both information on $m_t$ and on the value of certain other
hadronic matrix elements.  It is a vexing fact that the experimental
errors on all the measured parameters which are needed for testing the
CKM paradigm, are {\bf much less} than the corresponding theoretical
uncertainties which enter in the analysis.  For instance, the
experimental errors on $\epsilon$ and on the $B_d - {\bar{B}}_d$ mixing
parameter $x_d$ are, respectively, of order $1\%$ and $10\%$.  On the
other hand, the theoretical uncertainty which enters when one tries to
compare these parameters with the predictions of the CKM paradigm is of
order $50\%$!

It has become traditional to present the result of a CKM analysis of the
data as contour plots in the $\rho - \eta$ plane, as a function of the
top quark mass $m_t$ \cite{CKM analysis}.  Mostly because of the above
mentioned theoretical uncertainties, the measured values of $\epsilon$
and $x_d$ will map an allowed region in this plane.  For fixed $m_t$,
the theoretical uncertainty in $\epsilon$ arises from the uncertainty in
the value of $A$, as well as from a poor knowledge of the matrix element
of the $\Delta S = 2$ quark operator
\be
0_{\Delta S = 2} = (\bar{d} \gamma_{\mu} (1 - \gamma_5) s) (\bar{d}
\gamma^{\mu} ( 1 - \gamma_5) s)
\ee
between $K^0$ and ${\bar{K}}^0$ states.  The most reliable estimates for
$\vert V_{cb}\vert$, coming from the study of inclusive leptonic decays
\cite{AF}, as well as from the study of the exclusive decay ${\bar{B}}^0
\to D^{\ast +} \ell^- {\bar{v}}_{\ell}$ at zero recoil using heavy quark
techniques, \cite{Neubert} determine $A$ to a $10\%$ accuracy.  In what
follows, I shall use
\be
A = 0.9 \pm 0.1~~,
\ee
corresponding to $\vert V_{cb} \vert = 0.043 \pm 0.005$.

The hadronic matrix elements uncertanty in $\epsilon$ is usually
characterized by giving a value for the parameter $B_K$, which details
the ratio of the matrix element of $0_{\Delta S = 2}$ to that obtained
by using vacuum insertion.  The best estimates for $B_K$, coming from
lattice gauge theory computations \cite{Martinelli}, give
\be
B_K = 0.8 \pm 0.2~~.
\ee
The predicted value for the $B_d - {\bar{B}}_d$ mixing parameter $x_d$,
for fixed value of $m_t$, also depends on knowning $A$, but it requires
in addition, some knowledge of the $B_d$ decay constant $f_{B_{d}}$
defined by \footnote{Actually~\protect$x_d$ depends on the matrix element
of an operator \protect$0_{\Delta B = 2}$ analogous to~\protect$0_{\Delta
S = 2}$. This matrix element can be related to \protect$f_{B_{d}}^2
B_{B_{d}}$, with $B_{B_{d}}$ being the analogue of $B_K$ for the Kaon
case.  Because the $b$ quark is heavy, one expects the vacuum insertion
approximation to work very well, so that $B_{B_{d}} \simeq 1$.  In
addition, the formula for $x_d$ contains an overall factor of $\eta
\simeq 0.85$ multiplying $0_{\Delta B = 2}$ which accounts for short
distance QCD corrections to this operator \protect\cite{Gilman}.  In the text,
we report the value for \protect$\sqrt{B_{B_{d}} \eta} f_{B_{d}}$ which
is needed in the comparison of theory with experiment.}
\be
i f_{B_{d}} k^{\mu} = < 0 \vert \bar{d} \gamma_{\mu}
\gamma_5 b \vert B_d; k >~~.
\ee
The best value for this parameter, which follows from lattice QCD
computations, has also an error of about $10\%$.  One finds
\cite{Martinelli}
\be
\sqrt{B_{B_{d}} \eta} f_{B_{d}} = (200 \pm 35)~ MeV~~.
\ee

Theoretical formulas for $\vert \epsilon \vert$ and $x_d$
\cite{Inami}~\cite{Buras}  can be written in a handy approximate form
\cite{WW}, for $m_t > M_W$.  These formulas make it quite obvious what
is the source of the theoretical uncertainties and, given the
experimental value for $\vert \epsilon \vert$ and $x_d$, what is the
range allowed in the $\rho - \eta$ plane.  Using the values quoted for
$B_K$ and for $\sqrt{B_{B_{d}} \eta} f_{B_{d}}$, one has
\bea
\vert \epsilon \vert &\simeq & [ 2.7 \pm 0.7 ] \times 10^{-3} A^2 \eta \left\{
1 + \frac{4}{3} A^2 ( 1 - \rho) (\frac{m_t}{M_W})^{1.6}\right\}
\nonumber \\
x_d & \simeq&[0.44 \pm 0.15] A^2 [ \eta^2 + (1 - \rho)^2 ]
(\frac{m_t}{M_W})^{1.6}~~.
\eea
To the theoretical errors shown above, coming from our uncertain
knowledge of $B_K$ and $f_{B_{d}}$, one has to add the $20\%$
uncertainty present in $A^2$ to obtain,
from the experimental values for $\epsilon$ and $x_d
[ \vert \epsilon \vert = (2.268 \pm 0.023) \times 10^{-3}$ \cite{PDG};
$x_d = 0.64 \pm 0.08$ \cite{Cassel}] the allowed regions in the
$\rho - \eta$ plane. Figure 2 shows these regions for
the two cases:~$ m_t = 140~ GeV$ and $m_t = 180~GeV$.
\begin{figure}[h]
\vspace{8cm}
\caption[]{Allowed regions in the \protect$\rho - \eta$ plane
coming from the measurement of~\protect$\vert \epsilon \vert, x_d$ and
the ratio~\protect$\vert V_{ub}\vert/\vert V_{cb} \vert$.}
\end{figure}

In Figure 2, in addition to the allowed regions allowed by $\vert
\epsilon \vert$ and $x_d$, I indicated also the region in the $\rho -
\eta$ plane which is allowed by our present knowledge of the ratio of
$\vert V_{ub} \vert$ to $\vert V_{cb} \vert$.  A value for $\vert V_{ub}
\vert/\vert V_{cb} \vert$ fixes directly $\sigma$, or the value of
$\sqrt{\rho^2 + \eta^2}$. Including errors in $\vert V_{ub}
\vert/\vert V_{cb} \vert$ gives, therefore, the annular region centered
at $\rho = \eta = 0$ shown in Figure 2.  It is worthwhile also here to
discuss the source of the errors in $\vert V_{ub} \vert/\vert V_{cb}
\vert$ since, again, these are mostly due to theoretical uncertainties.

To extract $\vert V_{ub} \vert/\vert V_{cb}\vert$ from experiment one
studies the semileptonic decays of $B$ mesons $(B \to X \ell
v_{\ell})$ in a region of momentum of the emitted lepton $(p_{\ell} >
2.3~GeV)$ which insures kinematically that the hadronic states $X$ do
not contain a charmed quark.  That is, for $p_{\ell} > 2.3~GeV$ the
data should only measure decays in which the transition $b \to u$
occurred.  However, to extract a value of $\vert V_{ub}\vert$ from this
analysis is non trivial, since one must be able to estimate precisely
the hadronic matrix elements involved in the $ B \to X$ transition.
When one does this estimate by employing, as in the ACM model
\cite{ACM}, a
parton model - which is sensible in my mind, since one is summing over
all states $X$ - one gets a fairly large value for the
matrix element and hence a
rather small value for $\vert V_{ub} \vert/\vert V_{cb}\vert$.  On the
other hand, if one estimates the transition  $ B \to X$ by summing
only over some (assumed dominant) exclusive channels, as in the ISGW
model \cite{ISGW}, the strength of the transition is smaller and,
consequently, one deduces a larger value for $\vert V_{ub} \vert/\vert
V_{cb} \vert$.

Using only the more recent and more accurate data obtained by CLEO II,
Cassel \cite{Cassel} quotes the following values for $\vert V_{ub}\vert/\vert
V_{cb}\vert$ extracted, respectively, using the ACM model \cite{ACM} and the
ISGW model~\cite{ISGW}:
\bea
\bigg\vert \frac{V_{ub}}{V_{cb}}\bigg\vert &= 0.07 \pm 0.01
\leftrightarrow~ \sigma &= 0.32 \pm 0.06~~~{\rm ACM~ Model} \nonumber \\
\bigg\vert \frac{V_{ub}}{V_{cb}}\bigg\vert &= 0.11 \pm 0.02
\leftrightarrow~ \sigma &= 0.50 \pm 0.09~~~{\rm ISGW~Model}
\eea
The larger annulus in Figure 2 corresponds to taking the average of
these two results and somewhat expanding the errors by including other
model uncertainties \cite{Cassel}.  It corresponds to
\be
\bigg\vert \frac{V_{ub}}{V_{cb}} \bigg\vert =
0.085 \pm 0.045 \leftrightarrow \sigma = 0.39
\pm 0.21~~.
\ee
I have, however, also indicated in this figure the values of
$\sigma = \sqrt{\rho^2 + \eta^2}$ allowed if one extracted $\vert
V_{ub}\vert/\vert V_{cb}\vert$ from the data by using {\bf only} the ACM model.
As the figure makes clear,
it is rather important to resolve the theoretical controvery
surrounding the extraction of $\vert V_{ub} \vert/\vert V_{cb}\vert$
from experiment, as this
would considerably narrow the allowed region in
the $\rho - \eta$ plane. For example,
for $m_t = 140~GeV$, the overlap
region allowed by our present theoretical and experimental knowledge of
$\vert \epsilon \vert, x_d$ and $\vert V_{ub} \vert/\vert V_{cb}\vert$
is that shown in Fig. 3.  If one could trust the ACM model absolutely,
however, this region would get reduced to the rather narrow shaded band
shown in the figure.

\begin{figure}[h]
\vspace{6cm}
\caption[]{Allowed region in~\protect$\rho - \eta$ plane
for~\protect$m_t = 140~GeV$.  The shaded band is the result obtained by
relying only on the ACM model.}
\end{figure}

Unfortunately, even assuming $\eta$ to be in its most restricted range
$(\eta \simeq 0.2 - 0.3)$, is not sufficient to allow for a sharp
prediction for $\epsilon^{\prime}/\epsilon$, due to other theoretical
uncertainties arising
in estimating the hadronic matrix elements of operators
which contribute to $\epsilon^{\prime}$.  Nevertheless, considerable
progress has been made recently in trying to tackle this question,
notably by groups in Rome \cite{Ciuchini} and Munich \cite{Buchalla}
who have calculated the expectations for $\epsilon^{\prime}$ at next to
leading order and then tried to estimate the relevant matrix elements.
Because these calculations are highly technical, I will limit myself
here to give a more qualitative overview of the results obtained.

The ratio $\epsilon^{\prime}/\epsilon$ - which is essentially the same as $Re
{}~\epsilon^{\prime}/\epsilon$ - gets contribution from two kinds of
operators:  $\Delta I = 1/2$ operators and $\Delta I = 3/2$ operators.
The former contributions are induced by gluonic Penguins and thus are of
$0(\alpha_s)$.  However, since they enter in the amplitude $Im A_0$,
the $\Delta I = 1/2$ operators are affected by the whole $\Delta I
= 1/2$ suppression factor of $Re A_2/Re A_0 \simeq
1/20$ [c.f. Eq. (23)]. On the other hand,
the $\Delta I = 3/2$ contributions arise from
electroweak Penguin diagrams and thus are only of $0 (\alpha)$.
However, $Im A_2$ is measured relative to $Re A_2$ and so, effectively,
it is
not suppressed by the $\Delta I = 1/2$ factor of $Re
A_2/Re A_0$.  Furthermore, as first noted by Flynn and Randall
\cite{FR}, these contributions grow quadratically with $m_t$, while
those
of the gluonic Penguins only depends on $m_t$ as $\ell n~ m_t$.

In light of the above discussion, the structure of the result of the
calculations of $\epsilon^{\prime}/\epsilon$ can be written as
follows \cite{Ciuchini} \cite{Buchalla}:
\be
\frac{\epsilon^{\prime}}{\epsilon} = A^2 \eta \left\{
<2 \pi ; I = 0 \vert \sum_i C_i 0_i \vert K^0 > ( 1 - \Omega_I
)-< 2 \pi ; I = 2 \vert \sum_i {\tilde{C}}_i {\tilde{0}}_i \vert K^0
>\right\}
\ee
Here $0_i$ and ${\tilde{0}}_i$ are, respectively, $\Delta I = 1/2$ and
$\Delta I = 3/2$ operators and their coefficients $C_i$ and
${\tilde{C}}_i$ have the characteristic dependence on $\alpha_s \ell n~
m_t$ and $\alpha~ m_t^2$ alluded to above.  $\Omega_I$ is a correction to
the $\Delta I = 1/2$ contribution, which arises as a result of isospin
violation through $\pi^0 - \eta$ mixing \cite{BG} and is estimated to be
$\Omega_I = 0.25 \pm 0.10$.  Note also
in the above the characteristic
CKM dependence of $\epsilon^{\prime}$ - for a fixed given $\epsilon$ -
on the CKM parameters $A^2 \eta$.  Thus, even if the hadronic matrix
elements were perfectly known, present uncertainties in $A$ and $\eta$
would give about a $50\%$ uncertainty in $\epsilon^{\prime}/\epsilon$ - a bit
less if one could restrict $\eta$ to the ACM range.

It is difficult to extract directly from the work of the Rome
\cite{Ciuchini} and Munich \cite{Buchalla} groups a value for the
coefficient of $A^2 \eta$, typifying the hadronic uncertainty in
$\epsilon^{\prime}/\epsilon$.  Nevertheless, from these papers, more to
get a feeling for the expectatins than as a hard and fast result, I
infer the following.  For moderate $m_t$ - say $m_t = 140~GeV$ - gluonic
Penguins dominate.  Here the uncertainty in the matrix elements is more
under control, perhaps being only of order $30\%$.  A representative
prediction for $m_t$ in this range appears to be
\be
\frac{\epsilon^{\prime}}{\epsilon} = (11 \pm 4) \times 10^{-4} A^2
\eta~~~(m_t = 140~ GeV)~~.
\ee
For larger $m_t$ values $(m_t \simeq 200~ GeV)$ electroweak Penguins begin
to be important and they tend to cancel the contributions of the gluonic
Penguins.  The error in the matrix element estimation remains similar
in magnitude, but the central value for the overall
contribution is considerably reduced.  A representative prediction for
$m_t = 200~GeV$ is, perhaps,
\be
\frac{\epsilon^{\prime}}{\epsilon} = (3 \pm 4 ) \times 10^{-4} A^2 \eta
{}~~~(m_t = 200~GeV)~~.
\ee

If one takes the above numbers at face value, one sees that, with the
present range of $\eta$ allowed by the information on $\vert \epsilon
\vert, x_d$ and $\vert V_{ub}\vert/\vert  V_{cb} \vert$, the CKM
paradigm tends to favor rather small values for
$\epsilon^{\prime}/\epsilon$.  Typically, perhaps,
$\epsilon^{\prime}/\epsilon \simeq 4 \times 10^{-4}$, with a theory
error probably of the same order!  Such small values for
$\epsilon^{\prime}/\epsilon$ are perfectly compatible with the results
obtained by the E731 collaboration \cite{E731}, but are a bit difficult
to reconcile with the results of NA31 \cite{NA31}.

\section{Novel Tests of CPT at DAFNE and Fermilab}

As we saw earlier, present tests of CPT in the neutral Kaon system
involve in all cases {\bf differences} of CPT violating parameters.
Although I believe that cancellation among these parameters is unlikely,
one should soon be able to clarify this situation, with
experiments at DAFNE.  DAFNE, the high luminosity Phi Factory being
build at Frascati, may eventually reach a luminosity ${\cal{L}} =
10~^{33} cm^{-2} ~ sec^{-1}$, producing over $10^{10}$ correlated $K_L -
K_S$ pairs from $\Phi$ decay.  The possibility of studying such large
samples
of decays, as well as the particular features inherent from the way
these states are produced, makes DAFNE a very interesting machine to
further probe CP and CPT violation in the neutral Kaon sector.  As far
as CP goes, the KLOE detector at DAFNE should eventually be able to make
a measurement of $\epsilon^{\prime}/\epsilon$ competitive with what is
expected from the next round of experiments at CERN and Fermilab -
namely a measurement where the error on $\epsilon^{\prime}/\epsilon$ is
of the order of a few parts in $10^{-4}$.  However, it is in the realm
of CPT tests that DAFNE has a unique niche.

CPT can be further probed at DAFNE essentially because one can study
there, in addition to $K_L$ decays, also decays of the $K_S$. To test CPT
at DAFNE one can either:
\begin{description}
\item{i.)} use $K_L$ decays as a tag to study $K_S$ decays
\end{description}
or
\begin{description}
\item{ii.)} use the Phi factory directly as a $K^0 - {\bar{K}}^0$
interferometer.
\end{description}
In either case, one can make a direct measurement of $Re~ \delta_K$ and
from this knowledge then reconstruct all the individual CPT violating
parameters.  Let me briefly discuss how this can be accomplished for both
of the above methods. After doing so, I shall
summarize the results of an in depth
study~\cite{Buchanan}~which tried to estimate
the statistical accuracy with which one could measure CPT violating
parameters in a high luminosity Phi factory.

In addition to the semileptonic asymmetry for $K_L$ decays, at DAFNE one
will be able to measure, for the first time, the semileptonic asymmetry
in $K_S$ decays.  If CPT is conserved both of these asymmetries measure the
same CP violating parameter, $ 2 Re~ \epsilon_K$.  However, if CPT is
violated these asymetries will be different.  A simple calculation
\cite{Buchana
n}
gives
\bea
A_{K_{L}} = &2 Re~ \epsilon_K + [ 2 \frac{Re~ b}{Re~ a} + 2 Re \delta_K
]\nonumber \\
A_{K_{S}}  = &2 Re~ \epsilon_K + [2 \frac{Re~ b}{Re~ a} - 2 Re \delta_K]
\eea
Thus the difference between $A_{K_{L}}$ and $A_{K_{S}}$ provides a
direct measure of $Re~ \delta_K$.  For an integrated luminosity of $\int
{\cal{L}} d t = 10^{40}~cm^{-2}$, one should be able to obtain through
this comparison a measurement of $Re~ \delta_K$ to a statistical accuracy
of $0 (10^{-4})$\cite{Buchanan}.

One can also measure $Re~ \delta_K$, as well as $Im~ \delta_K$ (although
this is already known from $\epsilon_{c\not{p}t}$, by using the Phi
factory as a quantum interferometer \cite{DHR}.  The initial state in the Phi
factory, coming from the decay of the $\Phi$ into $K_L$  and $K_S$, is a
correlated superposition of $K_S$ and $K_L$ states. Taking the $\Phi$ to
be at rest, one has
\be
\vert \Phi > = \frac{1}{\sqrt{2}} \left\{ \vert K_S ( \vec{p}) > \vert
K_L (-\vec{p} ) > -\vert K_S (- \vec{p}) > \vert
K_L (\vec{p} ) > \right\}~~.
\ee
By measuring the relative time decay probability for observing the decay
by-products of the $K_L/K_S$ states into final states $f_1$ and $f_2$,
one translates this initial state correlation into a final state
interference pattern, whose precise shape will yield information on
possible CP and CPT violating parameters in the system.

Let the ratio of decay amplitudes of $K_L$ and $K_S$ into a final state
$f_i$ be denoted, as before, by
\be
\eta_i = \frac{A (K_L \to f_i)}{A(K_S \to f_i)} = \vert \eta_i \vert
e^{i \phi i}~~~.
\ee
Then a simple calculation \cite{Buchanan} gives the following
expression for the relative time decay probability for observing the
states $f_1$ and $f_2$, which were produced at times $t_1$ and $t_2$:
\begin{figure}[h]
\vspace{6cm}
\caption[]{Plot of the relative time decay probability for double
semileptonic decay versus $\Delta t$ is units of \protect$\tau_S$.  From
\protect\cite{Buchanan}.}
\end{figure}

\bea
I (f_1, f_2; \Delta t = t_1 - t_2) &=& \frac{1}{2} \int^{\infty}_{\Delta
t} d (t_1 + t_2 ) \vert < f_1 f_2 \vert T \vert \Phi > \vert^2 \nonumber
\\
&=&  {\rm const} \bigg\{\vert \eta_1 \vert^2 e^{-\Gamma_L \Delta t} + \vert
\eta_2 \vert^2 e^{- \Gamma_S \Delta t}\nonumber \\
&& 2 \vert \eta_1 \vert \vert \eta_2 \vert e^{- \frac{1}{2} (\Gamma_S +
\Gamma_L) \Delta t} \cos (\Delta m \Delta t + \phi_2 - \phi_1 )
\bigg\}~~.\nonumber \\
\eea
The above formula applies for $\Delta t > 0$.  If $\Delta t < 0$, then the
roles of $\Gamma_L$ and $\Gamma_S$ get interchanged.  Because of this, in
general, the pattern of the relative time decay probability is not
symmetric between positive and negative $\Delta t$.

To test CPT one can study the relative time decay probability pattern for
the case in which the final states $f_1$ and $f_2$ correspond to
semileptonic decays, e.g. $f_1 = \pi^- \ell^+ \nu_{\ell}$ and $f_2 = \pi^+
\ell^-{\bar{\nu}}_{\ell}$. In this case, one has \cite{Buchanan}
\be
\vert \eta_1 \vert^2 = 1 - 4 Re~ \delta_K~~;~~ \vert \eta_2 \vert^2 = 1 +
4 Re~ \delta_K
\ee
and
\be
\phi_1 - \phi_2 = \pi - 4 Im~ \delta_K~~.
\ee
As a result, there is an asymmetry in the decay of $I (f_1, f_2; \Delta
t)$ for $\Delta t >> \tau_S$ relative to $\Delta t < < - \tau_S$, with
the former decay being proportional to $(1 - 4 Re \delta_K) e^{-
\Gamma_L \Delta t}$ and the latter being proportional to $( 1 + 4 Re~
\delta_K) e^{- \Gamma_L \vert \Delta t \vert}$.  Fig. 4 shows the shape
of the expected relative time decay probability for this case.  The
interference dips near $\vert \Delta t \vert \simeq \tau_S$ are a measure
of $Im~ \delta_K$.
In Table 2, I show the results of the
comprehensive study of Buchanan et al \cite{Buchanan} on the
expected statistical accurary with which one can hope to measure, with
an integrated luminosity $\int {\cal{L}} dt = 10^{40}~cm^{-2}$, each of
the CPT violating parameters discussed earlier .  Note that none of
these parameters, except $Im \delta_K$, are measured individually at
present.
One sees from the above table that the results for each of the {\bf
individual parameters} at a high luminosity Phi factory should be of
comparable in
accuracy to present results on {\bf parameter
differences}.
\begin{table}
\caption{Accuracy expected for CPT violating parameters in a high
luminosity Phi factory.  From \protect\cite{Buchanan}}
\vspace*{.2cm}
\begin{center}
\begin{tabular}{|l|l|} \hline
${\rm Parameter}$ &${\rm Expected~ Error}$ \\
\hline
$Re~ \delta_K$ & $\pm 0.7 \times 10^{-4}$ \\
$Im~ \delta_K$ & $\pm 1.8 \times 10^{-4}$ \\
$Re~ b/ Re~ a$ & $\pm 1.9 \times 10^{-4}$\\
$Re B_0/Re A_0$ & $\pm 2.0 \times 10^{-4}$ \\
$Re B_2/Re A_2$ & $\pm 2.2 \times 10^{-4}$ \\ \hline
\end{tabular}
\end{center}
\end{table}

Clearly, although current data in the $K^0 - {\bar{K}}^0$ system is
consistent with CPT conservation, one will have to await the results
from DAFNE for really unambiguous tests.  This is perhaps best
illustrated by means of Figure 5.  Present measurements on the
difference between $\phi_{\epsilon}$ and $\phi_{sw}$ only tell us how
large $\epsilon_{c\not{p}t}$ is, but not how large individually are $Re
B_0/Re A_0$ and $Re \delta_K$.  At DAFNE, one will be able to measure
$Re \delta_K$ separately and thus deduce an unambiguous value for the
$K^0 - {\bar{K}}^0$ mass difference.  Although this mass difference
relative to $\Delta m$ will only be measured to the $10^{-4} $ level, on
absolute grounds one will measure
\be
\frac{m_{K^0} - m_{{\bar{K}}^0}} {m_{K^0}} \sim 0 (10^{ -18})
\ee

\begin{figure}[h]
\vspace{6cm}
\caption[]{Measurements of various CPT violating parameters possible in
a Phi factory.}
\end{figure}

\noindent This is a very interesting measurement, if CPT violating effects have
anything to do with the Planck Scale, $M_P \sim 10^{19}~GeV$.  Indeed,
if CPT violation occurs linearly in $M_P^{-1}$ \cite{Ellis}
one would expect
\be
m_{K^0} - m_{{\bar{K}}^0} \simeq m_{K^0} (\frac{m_{K^0}}{M_P}) \simeq
10^{-19} m_K
\ee
and one would be probing in the right parameter range at DAFNE!

Amazingly, this same range will be also probed by a forthcoming Fermilab
experiment which is set to measure the antiproton lifetime.  I would
like to conclude this talk by briefly discussing this experiment.  CPT
conservation requires that the lifetimes of particles and antiparticles
be the same.  Thus, if CPT is valid, since there exist
very strong bounds on
the proton lifetime one expects that the antiproton
lifetime should also be very long:
\be
\tau_{\bar{p}} = \tau_p \geq 10^{32}~{\rm years}~~.
\ee
However, at present, the most stringent direct bound on $\tau_{\bar{p}}$
is not even at the level of one year, being deduced from being able to
store successfully about $10^3~ \bar{p}$ in an ion trap.  One has
\cite{Gabrielse}
\be
\tau_{\bar{p}} < 3.4~ {\rm months}.
\ee

The above direct limit for the antiproton lifetime will be considerably
improved in the near future by the APEX experiment at Fermilab
\cite{APEX}. This experiment will search for $\bar{p}$ decays in the
Fermilab accumulator ring in a specially constructed vacuum tank,
designed to reduce the beam - gas background. A test of the APEX
experimental concept has already been performed, which has demonstrated
that the beam-gas background is understood.  In fact from this test one
can already set a much better (preliminary) limit \cite{Mueller},
\be
\tau_{\bar{p}} > 440~ {\rm years} B ( \bar{p} \to e^- \pi^0)~~,
\ee
on the antiproton lifetime than that from \cite{Gabrielse}.

The APEX experiments aims at reaching a limit for $\tau_{\bar{p}}$ of
the order $\tau_{\bar{p}} \geq 10^6 - 10^8$ years. Such a limit is
potentially interesting if CPT violating effects scale linearly with
$M^{-1}_P$. Indeed, imagine writing, for instance, the amplitude for the
decay
$\bar{p} \to e^- \pi^0$ as
\be
A (\bar{p} \to e^- \pi^0) = A(p \to e^+ \pi^0) + C (\frac{m_p}{M_P})^n~,
\ee
where the second term above represents possible CPT violating
contributions scaling as $M_P^{-n}$.  If $n = 1$, then the proton decay
amplitude in the above is negligible, and one has
\be
\tau_{\bar{p}} = \frac{5 \times 10^{-31}}{C^2} M^2_P~(GeV)~ BR
(\bar{p} \to e^- \pi^0)~ {\rm years}
\ee
With $C \sim 0(1)$ one sees that, if $n = 1$, then the $\bar{p}$
lifetime range probed by the APEX experiment could begin to be
interesting!  Of course, all of this is extremely speculative.  Indeed,
if one thinks of CPT violating amplitudes at the quark and lepton level,
purely on dimensional grounds one would expect $n$ to be 2 rather than 1
\cite{Ross}.

\section{Concluding Remarks}

I hope the above discussion has made clear that
present day experimental results on CP violation and on tests of CPT are
tantalizingly close to answering two very interesting and probing
questions.  Namely:
\begin{description}
\item{i.)} is there a $\Delta S = 1$ violation of CP, as expected in the CKM
paradigm?
\end{description}
and
\begin{description}
\item{ii.)} is CPT conserved to $0 \left( m_{\rm hadron}/M_P \right)$?
\end{description}
What is exciting is that
it is likely that one will get an answer to both of
these questions
rather soon.  Forthcoming experiments at CERN (NA48) and at Fermilab
(E832) are likely to resolve the $\epsilon^{\prime}/\epsilon$ issue once
and for all.  The KLOE collaboration at DAFNE has also the potential to
help resolve the present controvery regarding this parameter ratio.
Furthermore, KLOE as well as the
APEX experiment at Fermilab will provide important information regarding
CPT, at the level where logically there may be surprises.  Interesting
days are ahead!

\section{Acknowledgements}

This work was supported in part by the Department of Energy, under the
grant FG03-91ER 40662 Task C


\begin{thebibliography}{999}
\bibitem{PDG} Particle Data Group, K. Hikasa et al, Phy. Rev. D45, S1
(1992), D46, 5210(E) (1992).
\bibitem{CKM} N. Cabibbo, Phys. Rev. Lett. 10, 531 (1963); M. Kobayashi
and T. Maskawa, Prog. Theo. Phys. 49, 552 (1973)
\bibitem{CPT} J. Schwinger, Phys., Rev {\bf 82}, 914 (1951); G. Luders,
Dansk, Mat. Fys. Medd. {\bf 28,} 17 (1954); W., Pauli in {\it Niels
Bohr and the Development of Physics} (Pergamon Press, New York, 1955).
For  a recent discussion, see R. H. Dalitz, Nucl. Phys. B (Proc. Suppl.)
{\bf 24A}, 3 (1991).
\bibitem{Dib} C.0. Dib and R. D. Peccei, Phys. Rev. D 46, 2265 (1992)
\bibitem{Lee}T. D. Lee and C. S. Wu, Ann. Rev. Nucl. Sci 16, 471 (1966)
\bibitem{Ellis} T. Ellis, N. Mavromatos and D. Nanopolous, CERN
TH6755/92, to
be published in Nucl. Phys. B.
\bibitem{Buchanan} C. D. Buchanan, R. Cousin, C. Dib, R. D. Peccei and
J. Quackenbush, Phys. Rev. D45, 4088 (1992)
\bibitem{NA31} G. Barr in Proceedings of the Joint International Lepton
Photon Symposium and Europhysics Conference on HEP, S. Hegardy, K.
Potter and E. Quercigh, eds. Vol. 1 (1992) 179, (World Scientific,
Singapore, 1992); R. Carosi et al., Phys.
Lett. B127, 303~(1990)

\bibitem{E731} L. K. Gibbons et al, Phys. Rev. Lett. 70, 1199 (1993);
Phys. Rev. Lett. 70, 1203 (1993)
\bibitem{Cronin} J. W. Cronin, Rev. Mod. Phys. 53, 373, (1981)
\bibitem{Phase} L. K. Gibbons et al, Phys. Rev. Lett. 70, 1199 (1993)
\bibitem{G} C. Geweniger et al, Phys. Lett 52B, 119 (1974)
\bibitem{C} W. C. Carithers et al, Phys. Rev. Lett. 34, 1244 (1975)
\bibitem{GM} J. Gasser and U. G. Meissner, Phys. Lett 258B, 219, (1991)
\bibitem{Ochs} N. Ochs, MPI-Ph/PL 91-35 preprint
\bibitem{Wolf} L. Wolfenstein, Phys. Rev. Lett 51, 1945 (1983)
\bibitem{CKM analysis} P. Krawczyk, D. London, R. D. Peccei and H.
Steger, Nucl. Phys. B307, 19~(1988); C. O. Dib, I. Dunietz, F.
J. Gilman, and Y. Nir, Phys. Rev. D41,  1522~(1990); C. S. Kim,
J. L. Rosner and C. P. Yuan,  Phys. Rev.  D42,  96~(1990); G.
Belanger and C. Q. Geng,  Phys. Rev. D43,  l40~(1991); M.
Lusignoli, L. Maiani, G. Martinelli and L. Reina  Nucl. Phys.
B369, 139~(1992); J. L. Rosner  J. Phys. G: Nucl. Part. Phys~
18, 1575 (1992)
\bibitem{AF} See for example, S. L. Stone in {\it B. decays}~S. L. Stone
ed. (World Scientific, Singapore 1992)
\bibitem{Neubert} M. Neubert, Phys. Lett 264B, 455 (1991)
\bibitem{Gilman} J. S. Hagelin, Nucl. Phys. B193, 123 (1981)
\bibitem{Martinelli} G. Martinelli, Nucl. Phys. B. (Proc. Suppl) 26, 31
(1992)
\bibitem{Martinelli 2} G. Martinelli, Proceedings of the PASCOS/TEXAS
Symposium, Berkeley, California, Dec. 1992
\bibitem{Inami} T. Inami and C. S. Lim, Prog. Theo. Phys. 65, 1297
(1981)
\bibitem{Buras} A. J. Buras, W. Slominski and H. Steger, Nucl. Phys.
B238, 529 (1984)
\bibitem{WW} B. Winstein and L. Wolfenstein, CMU-DOE ER/40682-24,
submitted to Rev. Mod. Phys.
\bibitem{Cassel} D. Cassel, Proceedings of the 1992 DPF Conference,
Fermilab, Batavia, Illinois
\bibitem{ACM} G. Altarelli et al, Nucl. Phys. B207, 365 (1982)
\bibitem{ISGW} N. Isgur, D. Scora, B. Grinstein and M. B. Wise, Phys.
Rev. D39 799 (1989)
\bibitem{Ciuchini} M. Ciuchini, E. Franco, G. Martinelli and L. Reina,
Rome preprint 911, 1992
\bibitem{Buchalla} A. J. Buras, M. Jamin and M. E. Lautenbacher,
MPI-Ph/93-11
\bibitem{FR} J. Flynn and L. Randall, Phys. Lett. B216, 221 (1989);
Nucl. Phys. B326, 31 (1989)
\bibitem{BG} A. J. Buras and J. M. Gerard, Phys. Lett B 192, 156 (1987)
\bibitem{DHR} H. J. Lipkin, Phys. Rev. 176 1785 (1968); I. Dunietz, J.
Hauser and J. Rosner, Phys. Rev. D35, 2166 (1987); J. Bernabeu, F. J.
Botella and J. Roldan, Phys. Lett. B211, 226 (1980)
\bibitem{Gabrielse} G. Gabrielse et al, Phys. Rev. Lett. 65, 1317 (1990)
\bibitem{APEX} S. Geer et al, APEX proposal
\bibitem{Mueller} Thomas Mueller, private communication
\bibitem{Ross} Graham Ross, private communication
\end{thebibliography}
\end{document}